\documentclass{article}
\usepackage[table,xcdraw]{xcolor}
\usepackage{spconf,amsmath,graphicx}
\usepackage{url}
\usepackage{array}
\usepackage{amssymb}
\usepackage{multirow}
\usepackage{caption}
\usepackage{subcaption}
\usepackage{color}
\usepackage{tikz}
\usepackage{booktabs}
\usepackage{enumitem}
\usepackage{multirow}
\usepackage{multicol}
\usepackage{pifont}
\usepackage{tablefootnote}
\usepackage{caption}
\usepackage{subcaption}
\usepackage{cite}

\newcommand\blfootnote[1]{%
  \begingroup
  \renewcommand\thefootnote{}\footnote{#1}%
  \addtocounter{footnote}{-1}%
  \endgroup
}

\usepackage[pagebackref,bookmarks=false,hidelinks]{hyperref} 
\renewcommand*{\backref}[1]{} \renewcommand*{\backrefalt}[4]{{\footnotesize(\ifcase #1 Not cited.\or Cited on page~#2.\else Cited on pages #2.\fi)}}

\title{Large-scale Contrastive Language-Audio Pretraining with \\ Feature Fusion and Keyword-to-Caption Augmentation}

\name{
\parbox{\linewidth}{\centering
Yusong Wu~$^{1*}$, Ke Chen~$^{2*}$, Tianyu Zhang~$^{1*}$, Yuchen Hui~$^{1,3*}$, Marianna Nezhurina~$^{1,3}$ \\ Taylor Berg-Kirkpatrick~$^{2}$, Shlomo Dubnov~$^{2}$
}}
\address{
         $^1$Mila, Quebec Artificial Intelligence Institute, Université de Montréal \\
         $^2$University of California San Diego ~~~ $^3$LAION \\
}

\begin{document}
\ninept 

\maketitle

\begin{abstract}
Contrastive learning has shown remarkable success in the field of multimodal representation learning.
In this paper, we propose a pipeline of contrastive language-audio pretraining to develop an audio representation by combining audio data with natural language descriptions. To accomplish this target, we first release \texttt{LAION-Audio-630K}, a large collection of 633,526 audio-text pairs from different data sources. Second, we construct a contrastive language-audio pretraining model by considering different audio encoders and text encoders. We incorporate the feature fusion mechanism and keyword-to-caption augmentation into the model design to further enable the model to process audio inputs of variable lengths and enhance the performance\blfootnote{*The first four authors have equal contribution. Marianna Nezhurina was not listed as the author in the paper of the ICASSP 2023 proceedings. We acknowledge her contribution to the LAION-Audio-630K collection.}. Third, we perform comprehensive experiments to evaluate our model across three tasks: text-to-audio retrieval, zero-shot audio classification, and supervised audio classification. The results demonstrate that our model achieves superior performance in text-to-audio retrieval task. In audio classification tasks, the model achieves state-of-the-art performance in the zero-shot setting and is able to obtain performance comparable to models' results in the non-zero-shot setting. \texttt{LAION-Audio-630K}\footnote{Dataset: \href{https://github.com/LAION-AI/audio-dataset/}{https://github.com/LAION-AI/audio-dataset/}} and the proposed model\footnote{Model: \href{https://github.com/LAION-AI/CLAP}{https://github.com/LAION-AI/CLAP}} are both available to the public. 
\end{abstract}
\begin{keywords}
Contrastive Learning, Representation Learning, Text-to-Audio Retrieval, Audio Classification, Audio Dataset
\end{keywords}

\section{Introduction}
Audio is one of the most common information types in the world alongside text and image data.
However, different audio tasks typically require finely-annotated data, which limits the amount of available audio data due to the labor-intensive collection procedure. 
Consequently, designing an effective audio representation for many audio tasks without requiring a lot of supervision remains a challenge.

The contrastive learning paradigm is a successful solution for training a model on large-scale noisy data collected from internet. 
The recently proposed Contrastive Language-Image Pretraining (CLIP) \cite{clip} learns the correspondence between text and image by projecting them into a shared latent space. The training is conducted by regarding the ground-truth image-text pair as the positive sample and left as negative. 
In contrast to training on unimodal data, CLIP is not constrained by data annotation and shows great robustness by achieving high accuracy in a zero-shot setting on out-of-domain variations of ImageNet dataset \cite{imagenet}. 
Additionally, CLIP shows great success in downstream tasks such as text-to-image retrieval and text-guided captioning. 
Similar to vision, audio and natural languages also contain overlapping information.
In audio event classification task, for instance, some text descriptions of an event can be mapped to the corresponding audio.
These text descriptions share a similar meaning that could be learned together with the related audio to form an audio representation of crossmodal information. Additionally, training such a model requires simply paired audio and text data, which is easy to collect.

Several recent studies \cite{audioclip,clap,clap-retrieval,yuke-dcase,mmt,ml-act,wav2clip} have presented the prototype of the contrastive language-audio pretraining model for the text-to-audio retrieval task. 
\cite{yuke-dcase} utilizes Pretrained Audio Neural Network (PANN) \cite{pann} as the audio encoder, BERT \cite{bert} as the text encoder, and several loss functions to evaluate the text-to-audio retrieval performance.
\cite{clap-retrieval} further ensemble HTSAT \cite{hts-at} and RoBERTa \cite{roberta} into the encoder list to further enhance performance.
Then, \cite{clap} investigates the effectiveness of the learned representation in the downstream task of audio classification.
Some other studies, such as AudioClip \cite{audioclip} and WaveCLIP \cite{wav2clip}, focus more on the contrastive image-audio (or image-audio-language) pretraining model. 
All these models show great potential for contrastive learning in the audio domain.

Nonetheless, current studies have not shown the full strength of the language-audio contrastive learning.
First, the models mentioned above are trained on relatively small datasets, showing that large-scale data collection and augmentation for training are needed. Second, prior work lacks a full investigation of selections and hyperparameter settings of audio/text encoders, which is essential for determining the basic contrastive language-audio architecture.
Third, the model struggles to accommodate varied audio lengths, particularly for the transformer-based audio encoder.
There should be a solution to handle audio inputs of variable-length.
Finally, the majority of language-audio model studies focuses solely on text-to-audio retrieval without assessing their audio representations in downstream tasks. As a representation model, we expect more discoveries of its generalization ability to more downstream tasks.

In this paper, we make contributions to improve the dataset, model design and the experiment setting from above concerns:
\begin{itemize}[leftmargin=*]
    \item We release LAION-Audio-630K, currently the largest public audio caption dataset of 633,526 audio-text pairs. To facilitate the learning process, we employ the keyword-to-caption model to augment labels of AudioSet \cite{audioset} into corresponding captions. This dataset can also contribute to other audio tasks.  
    
    \item We construct a pipeline of contrastive language-audio pretraining. Two audio encoders and three text encoders are selected for testing. We employ feature fusion mechanisms to enhance the performance and enable our model to handle variable-length inputs.

    \item We conduct comprehensive experiments on the model, including the text-to-audio retrieval task, as well as zero-shot and supervised audio classification downstream tasks. We demonstrate that scaling of the dataset, keyword-to-caption augmentation, and feature fusion can improve the model's performance in different perspectives. It achieves the state-of-the-art (SOTA) in the text-to-audio retrieval and audio classification tasks, even comparable to the performance of supervised models.
\end{itemize}
We make both LAION-Audio-630K and the proposed model available to the public.

\section{LAION-Audio-630K and Training Dataset}
\begin{table}[t]
\centering
\resizebox{\columnwidth}{!}{
\begin{tabular}{lcc}
\toprule
Dataset & Pairs & Audio Durations (hrs)   \\
\midrule 
Clotho\cite{clotho} & 5,929 & 37.00  \\
SoundDescs\cite{audiotextbenchmark} & 32,979 & 1060.40 \\
AudioCaps\cite{audiocaps} & 52,904 & 144.94 \\
LAION-Audio-630K & 633,526 & 4325.39 \\
\bottomrule
\end{tabular}}
\caption{LAION-Audio-630K compared with existing datasets.}
\vspace{-0.8cm}
\label{tab:LAION-Audio-630K scale comparation}
\end{table}
\subsection{LAION-Audio-630K}\label{sec:dataset}

We collect LAION-Audio-630K, a large-scale audio-text dataset consisting of 633,526 pairs with the total duration of 4,325.39 hours. 
It contains audios of human activities, natural sounds and audio effects, consisting of 8 data sources from publicly available websites\footnote{Dataset details are appended at: \href{https://retrocirce.github.io/appendix/}{https://retrocirce.github.io/appendix/}}. 
We collect these datasets by downloading audios and relevant text descriptions. Based on our current knowledge, LAION-Audio-630K is the largest audio-text dataset publicly available and a magnitude larger than previous audio-text datasets as shown in Table~\ref{tab:LAION-Audio-630K scale comparation}.

\vspace{-0.3cm}
\subsection{Training Dataset} \label{sec:train_dataset}
To test how model performance will scale on different sizes and types of dataset, we use three training set setting in the paper, varying from small to large size. These settings employ three datasets: 1) \textbf{AudioCaps+Clotho} (\textbf{AC+CL})~\cite{audiocaps,clotho} contains about 55K training samples of audio-text pairs. 2) LAION-Audio-630K (\textbf{LA.}) consists of around 630K audio-text pairs. 
3) \textbf{Audioset}~\cite{audioset} consists of 1.9 million audio samples with only labels available for each sample. When processing these datasets, we exclude all overlapping data in evaluation sets. More details of the training datasets can be found at the online appendix.

\vspace{-0.3cm}
\subsection{Dataset Format and Preprocessing}
All audio files used in this work are preprocessed to mono channel at a sample rate of 48kHz in FLAC format. For datasets with only tags or labels available, we extend labels into captions using the template ``The sound of \texttt{label-1}, \texttt{label-2}, ..., and \texttt{label-n}" or the keyword-to-caption model (detail in section \ref{sec:ksa}). As a result, we can leverage more data into the training of the contrastive language-audio pretraining model. Combining all the datasets, we increase the total number of audio samples with text caption to 2.5 million.

\section{Model Architecture}
\begin{figure}[t]
    \centering
    \includegraphics[width=\columnwidth]{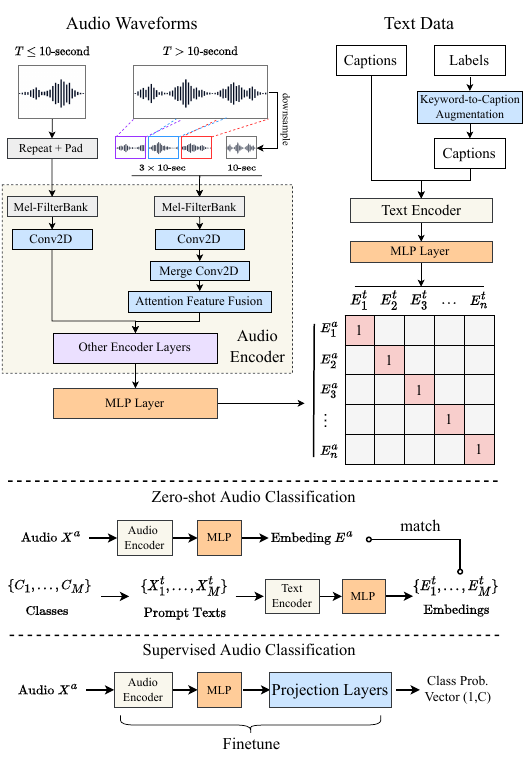}
    \caption{The architecture of our proposed model, including audio/text encoders, feature fusion, and keyword-to-caption augmentation.}
    \label{fig:audioclip-arch}
    \vspace{-0.5cm}
\end{figure}

\subsection{Contrastive Language-Audio Pretraining}
Figure \ref{fig:audioclip-arch} depicts the general architecture of our proposed contrastive language-audio encoder model. Similar to CLIP \cite{clip}, we have two encoders to separately process the input of audio data $X_i^a$ and text data $X_i^t$, where $(X_i^a, X_i^t)$ is one of audio-text pairs indexed by $i$. The audio embedding $E_i^{a}$ and the text embedding $E_i^{t}$ are respectively obtained by the audio encoder $f_{audio}(\cdot)$ and the text encoder $f_{text}(\cdot)$, with projection layers:
\begin{align}
    E_i^{a} &= MLP_{audio}(f_{audio}(X_i^a)) \\
    E_i^{t} &= MLP_{text}(f_{text}(X_i^t)) 
\end{align}
Where the audio/text projection layer is a 2-layer multilayer perceptron (MLP) with ReLU \cite{relu} as the activation function to map the encoder outputs into the same dimension $D$ (i.e., $E_i^{a}, E_i^{t} \in \mathbb{R}^D$).

The model is trained with the contrastive learning paradigm between the audio and text embeddings in pair, following the same loss function in \cite{clip}:
\begin{align}
\begin{split}
\resizebox{\columnwidth}{!}{
    $L = \frac{1}{2N} \sum_{i=1}^N (\log \frac{\exp(E_i^a \cdot E_i^t / \tau)}{\sum_{j=1}^N \exp(E_i^a \cdot E_j^t / \tau)} + \log \frac{\exp(E_i^t \cdot E_i^a / \tau)}{\sum_{j=1}^N \exp(E_i^t \cdot E_j^a / \tau)})$
}
\end{split}
\end{align}
Where $\tau$ is a learnable temperature parameter for scaling the loss. Two logarithmic terms consider either audio-to-text logits or text-to-audio logits. $N$ is usually the number of data, but during the training phase, $N$ is used as the batch size, as we cannot compute the whole matrix of all data but update the model by batch gradient descent. 

After we train the model, the embeddings $(E^a, E^b)$ can be used for different tasks as shown in Figure \ref{fig:audioclip-arch} and listed in the below subsection.

\subsection{Downstream Tasks in Inference Stage}
\noindent\textbf{Text-to-Audio Retrieval}~ The target audio embedding $E_p^a$ can find the nearest text embedding $E_q^t$ among $M$ texts $E^t=\{E_1^t, ..., E_M^t\}$ by the cosine similarity function, determining the best match.

\vspace{0.1cm}
\noindent\textbf{Zero-shot Audio Classification}~ For $M$ audio classes $C=\{C_1,...,C_M\}$, we can construct $M$ prompt texts $X^t=\{X_1^t,...,X_M^t\}$ (e.g., ``the sound of \texttt{class-name}"). For a given audio $X_p^a$, we determine the best match $X_q^t$ among $X^t$ by the cosine similarity function over their embeddings. One advantage of using the contrastive language-audio pretraining is that the categories of audio are unrestricted (i.e., zero-shot) since the model can convert the classification task into the text-to-audio retrieval task.

\vspace{0.1cm}
\noindent\textbf{Supervised Audio Classification}~ After training the model, for a given audio $X_p^a$, its embedding $E_p^a$ can be further mapped into a fixed-category classification task by adding a projection layer at the back and finetuning (i.e., the non-zero-shot setting).

\subsection{Audio Encoders and Text Encoders}\label{sec:at-encoder}
We select two models, PANN \cite{pann} and HTSAT \cite{hts-at}, to construct the audio encoder. PANN is a CNN-based \cite{cnn} audio classification model with 7 downsampling CNN blocks and 7 upsampling blocks. HTSAT is a transformer-based model with 4 groups of swin-transformer blocks \cite{swintransformer}, which achieves SOTAs on three audio classification datasets. For both of them, we use their penultimate layer's output, a $L$-dimension vector as the output sent to the projection MLP layer, where $L_{PANN}=2048$ and $L_{HTSAT}=768$.

We select three models, CLIP transformer \cite{clip} (text encoder of CLIP), BERT \cite{bert}, and RoBERTa \cite{roberta}, to construct the text encoder. The output dimension of text encoders is respectively $L_{CLIP}=512$. $L_{BERT}=768$, and $L_{RoBERTa}=768$. We apply both 2-layer MLPs with ReLU activation \cite{relu} to map both audio and text outputs into 512 dimensions, which is the size of audio/text representations when training with the contrastive learning paradigm.

\subsection{Feature Fusion for Variable-Length Audio}
Unlike RGB image data that can be resized to a unified resolution, audio has a nature of variable length. 
Conventionally, one would input the full audio into the audio encoder and take the average of per-frame or per-chunk audio embeddings as output (i.e., slice \& vote). However, the conventional method is computationally inefficient on long audio.
As shown in the left of Figure \ref{fig:audioclip-arch}, we train and inference on different lengths of audio inputs in constant computation time by combining both coarsely global and randomly sampled local information. For an audio in $T$ seconds and a fixed chunk duration $d=10$ seconds:
\begin{itemize}[leftmargin=*]
    \item $T \leq d$: we first repeat the input, then pad it with zero values. For example,  a $\text{3-second}$ input will be repeated as $3\times3=\text{9-second}$ and padded with $\text{1-second}$ zero values.
    \item $T > d$: we first downsample the input from $T$ to $d$-second as a global input. Then we randomly slice three $d$-second clips, respectively from the front $\frac{1}{3}$, middle $\frac{1}{3}$ and back $\frac{1}{3}$ of the input, as local inputs. We send these $4 \times d$ inputs into the first layer of audio encoder to get the initial features, then three local features will be further converted to one feature by another 2D-Convolution layer with 3-stride in the time axis. Finally, the local feature $X^a_{local}$ and the global feature $X^a_{global}$ will be fused as:
    \begin{align}
        X^a_{fusion}= \alpha X^a_{global}+(1-\alpha)X^a_{local}
    \end{align}
    Where $\alpha=f_{AFF}(X^a_{global},X^a_{local})$ is a factor obtained by attention feature fusion (AFF) \cite{aff}, a two-branch CNN model for learning the fusion factor of two inputs. Comparing with the ``slice \& vote" method, the feature fusion also saves the training time as we only process audio slices in the first few layers. 
\end{itemize}
\vspace{-0.4cm}
\subsection{Keyword-to-Caption Augmentation} \label{sec:ksa}
As mentioned in section \ref{sec:dataset}, some datasets contains reasonable labels or tags as keywords of the corresponding audios. As shown in the right of Figure \ref{fig:audioclip-arch}, we used a pre-trained language model T5 \cite{t5model} to make captions on top of these keywords. We also de-bias the output sentence as post-processing. For example, we replace ``woman" and ``man" with `person' as gender de-biasing. Due to the page limit, we provide examples of the augmentation in the online appendix. 

\vspace{-0.3cm}
\section{Experiments} \label{sec:exp}
\vspace{-0.2cm}
In this section, we conduct three experiments on our proposed model. First, we train with different audio and text encoders to find the best baseline combination. Then, we train our model on various dataset size, with the feature fusion and keyword-to-caption augmentation to verify the efficacy of the proposed methods.
For the first two experiments, we evaluate our model's performance via recall and mean average precision (mAP) on audio-to-text and text-to-audio retrieval. Lastly, we use the best model to conduct zero-shot and supervised audio classification experiments to evaluate the generalization ability to the downstream tasks.

\begin{table}[t]
\centering
\resizebox{0.9\columnwidth}{!}{
\begin{tabular}{lcccc}
\toprule
\multirow{2}{*}{Model} & \multicolumn{2}{c|}{AudioCaps (mAP@10)}                    & \multicolumn{2}{c}{Clotho (mAP@10)}   \\ \cline{2-5} 
                       & ~~~A$\rightarrow$T & \multicolumn{1}{c|}{T$\rightarrow$A} & ~~A$\rightarrow$T & T$\rightarrow$A \\ \hline
PANN+CLIP Trans.       & 4.7               & 11.7                                   & 1.9               & 4.4               \\
PANN+BERT              & 34.3              & 44.3                                   & 10.8              & 17.7              \\
PANN+RoBERTa           & 37.5              & 45.3                                   & 11.3              & 18.4              \\
HTSAT+CLIP Trans.      & 2.4               & 6.0                                    & 1.1               & 3.2               \\
HTSAT+BERT             & 43.7              & 49.2                                   & \textbf{13.8}     & \textbf{20.8}     \\
HTSAT+RoBERTa          & \textbf{45.7}     & \textbf{51.3}                          & \textbf{13.8}     & 20.4              \\ \hline
\end{tabular}
}
\caption{The text-to-audio retrieval result (mAP@10) of using different audio/text encoder on AudioCaps and Clotho.}
\vspace{-0.5cm}
\label{tab:exp-ta-abalation}
\end{table}

\begin{table*}[t]
\vspace{-0.2cm}
\resizebox{\textwidth}{!}{
\begin{tabular}{lccccccc|cccccc}
\hline\hline
\multicolumn{1}{c}{\multirow{3}{*}{Model}} &
  \multirow{3}{*}{Training Set} &
  \multicolumn{6}{c|}{AudioCaps  Eval.} &
  \multicolumn{6}{c}{Clotho Eval.} \\ \cline{3-14} 
\multicolumn{1}{c}{} &
   &
  \multicolumn{3}{c}{T-A Retrieval} &
  \multicolumn{3}{c|}{A-T Retrieval} &
  \multicolumn{3}{c}{T-A Retrieval} &
  \multicolumn{3}{c}{A-T Retrieval} \\ \cline{3-14} 
\multicolumn{1}{c}{}  &                                              & R@1   & R@5  & R@10 & R@1  & R@5  & R@10 & R@1  & R@5  & R@10 & R@1  & R@5  & R@10 \\ \hline
MMT \cite{mmt}                   & AudioCaps or Clotho                             & 36.1  & \textbf{72.0} & \textbf{84.5} & 39.6 & 76.8 & 86.7 & 6.7  & 21.6 & 33.2 & 7.0  & 22.7 & 34.6 \\
ML-ACT \cite{ml-act}                & AudioCaps or Clotho                             & 33.9  & 69.7 & 82.6 & 39.4 & 72.0 & 83.9 & 14.4 & 36.6 & 49.9 & 16.2 & 37.6 & 50.2 \\
CLAP-HTSAT \cite{clap-retrieval} &
  AudioCaps + Clotho + WT5K & 34.6 & 70.2 & 82.0 & 41.9 & 73.1 & 84.6 & 16.7 & 41.1 & 54.1 & 20.0 & 44.9 & 58.7 \\ \hline
HTSAT-RoBERTa    & AudioCaps + Clotho                              &  \textbf{36.7}  &  70.9    & 83.2 &  45.3 &  78.0  &  87.7  & 12.0 & 31.6  &  43.9    &   15.7   &  36.9    &  51.3    \\
HTSAT-RoBERTa    & AudioCaps + Clotho + LA.                       &  32.7    &  68.0  &  81.2   &  43.9   &  77.7    &  87.6    &  15.6    &   38.6  &  52.3    & 23.7    &  48.9   &  59.9    \\
HTSAT-RoBERTa (fusion)    & AudioCaps + Clotho + LA.                       &  36.2   &  70.3  &  82.5   &  45.0   &  76.7    &  88.0    &  \textbf{17.2}  &  \textbf{42.9}  &  \textbf{55.4}    & 24.2    &  \textbf{51.1}   &  \textbf{66.9}    \\
HTSAT-RoBERTa    & ACaps. + Clotho + LA. + AudioSet (template) &  34.7  &  70.5 &  83.2 &   45.3   &   79.5  &  89.2  &  16.4    &  39.0    &  51.0    &  21.8   &  44.6   &  60.1    \\ 
HTSAT-RoBERTa    & ACaps. + Clotho + LA. + AudioSet (K2C aug.) &  36.1  &  71.8  &  83.9 &   \textbf{46.8}   &   \textbf{82.9}   &  \textbf{90.7}  &  16.1    &  38.3    &  51.1    &  22.7   &  48.5   &  60.8    \\ 
HTSAT-RoBERTa (fusion)    & ACaps. + Clotho + LA. + AudioSet (K2C aug.) &  35.1  & 71.9   &  83.7 &   44.2   &  80.8    & 90.3  &  16.9    &  41.6    & 54.4     & \textbf{24.4}    & 49.3    &  65.7 \\\hline\hline 
\end{tabular}}
\vspace{-0.2cm}
\caption{The text-to-audio retrieval performance on AudioCaps and Clotho datasets, where ``LA." refers to LAION-Audio-630K, ``template" refers to the text prompting by templates, ``K2C aug." refers to the keyword-to-caption augmentation, and ``fusion"  refers to the feature fusion.}
\vspace{-0.5cm}
\label{tab:exp-t2a-retrieval}
\end{table*}

\vspace{-0.2cm}
\subsection{Hyperparameters and Training Details}
As mentioned in section \ref{sec:train_dataset}, we use AudioCaps, Clotho, LAION-Audio-630K, along with the additional dataset --- AudioSet by keyword-to-caption augmentation, to train our model. For the audio data, we use 10-second input length, 480 hop size, 1024 window size, 64 mel-bins to compute STFTs and mel-spectrograms. As the result, each input sent to the audio encoder is of the shape $(T=1024, F=64)$. For the text data, we tokenize the text with a maximum token length of 77. 

When training the model without the feature fusion, the audio longer than 10-second will be randomly chunked to a 10-second segment. During training, we use the Adam \cite{kingma2014adam} optimizer with $\beta_1=0.99$, $\beta_2=0.9$ with a warm-up \cite{goyal2017accurate} and cosine learning rate decay at a basic learning rate of $10^{-4}$. We train the model using a batch size of 768 on \textbf{AudioCaps+Clotho} dataset, 2304 on training dataset containing LAION-Audio-630K, and 4608 on training dataset containing \textbf{AudioSet}. We train the model for 45 epochs.

\vspace{-0.2cm}
\subsection{Text-to-Audio Retrieval}

\noindent \textbf{Audio and Text Encoders} We first conduct experiments to choose the best audio encoder and text encoder for the text-to-audio retrieval task. We combine two audio encoders with three text encoders in section~\ref{sec:at-encoder} where both are loaded from pretrained checkpoints as the same to \cite{mmt,ml-act,clap-retrieval}. In this experiment, we only train on AudioCaps and Clotho datasets ($\sim$55K data), and report the best mAP@10 on audio-to-text (A$\rightarrow$T) and text-to-audio (T$\rightarrow$A) perspectives.

According to the results in Table \ref{tab:exp-ta-abalation}, for audio encoder, HTSAT performs better than PANN combined with the RoBERTa or BERT text encoder. For the text encoder, RoBERTa achieves better performance than BERT while the CLIP transformer performs the extremely worst. This coincides with the choice of text encoder in previous works~\cite{ml-act,clap}. When further analyzing the loss convergence trends of CLIP transformer model, we find that RoBERTa is less over-fitting, while CLIP transformer is of high-over-fitting, thus resulting its low generalization performance. 

\vspace{0.05cm}
\noindent \textbf{Dataset Scale} Consequently, we apply HTSAT-RoBERTa as our best model setting to conduct the text-to-audio retrieval experiments as a comprehensive evaluation in Table~\ref{tab:exp-t2a-retrieval}. We adopt the same metrics in \cite{ml-act,mmt} to compute recall scores at different ranks in this task. In the training set, we gradually increase the scale of the dataset.
We find that scaling up the dataset from ``AudioCaps + Clotho" to ``LA." does not improve the result on AudioCaps evaluation set but gets better performance on Clotho evaluation set, which is similar to the comparison between MMT \cite{mmt} and CLAP-HTSAT \cite{clap-retrieval}. One reason is that AudioCaps contains audios similar to AudioSet on which the audio encoder's loaded checkpoint is pretrained. When the model receives more data from other sources, it increases its generalization but moves the distribution out of AudioSet data. Therefore, the performance on AudioCaps drops but that on Clotho increases a lot, demonstrating a trade-off of the model to keep the performance among different types of audios. 

\vspace{0.05cm}
\noindent \textbf{Keyword-to-Caption and Feature Fusion} When adding the feature fusion mechanism and keyword-to-caption augmentation to the model, we can observe that either of them improves the performance. The feature fusion is effective especially in Clotho dataset because it contains longer audio data ($>10\text{-second}$). When we add AudioSet into the training set with either template prompting or keyword-to-caption augmentation, we can see the performance increases again on AudioCaps while decreases on Clotho. This further confirms the trade-off performance between AudioCaps and Clotho datasets mentioned above. And the keyword-to-caption augmentation does bring in better performance than the simple template text prompting method on most metrics.

As the result, our best model outperforms previous methods on most metrics (mainly R@1=36.7\% on AudioCaps and R@1=18.2\% on Clotho) in the text-to-audio retrieval tasks. We show that training on large-scale datasets (LAION-Audio-630K and AudioSet with keyword-to caption augmentation), and feature fusion can effectively improve model performance.

\begin{table}[]
\resizebox{\columnwidth}{!}{
\begin{tabular}{lccccc}
\hline\hline
\multirow{3}{*}{Model} & \multicolumn{5}{c}{Audio Classification Dataset \& Setting}                                                          \\ \cline{2-6} 
                       & \multicolumn{1}{c|}{ESC-50} & \multicolumn{1}{c|}{US8K} & \multicolumn{2}{c|}{VGGSound}                     & FSD50K \\ \cline{2-6} 
                       & \multicolumn{1}{c|}{ZS.}     & \multicolumn{1}{c|}{ZS.}   & \multicolumn{1}{c|}{ZS.} & \multicolumn{1}{c|}{SV.} & SV.     \\ \hline
Wav2CLIP\cite{wav2clip}               & 41.4                        & 40.4                      & 10.0                    & 46.6                    & 43.1   \\
AudioClip\cite{audioclip}              & 69.4                        & 65.3                      & -                       & -                       & -      \\
Microsoft\cite{clap-retrieval}                   & 82.6                        & 73.2                      & -                       & -                       & 58.6   \\ \hline
CLAP                   & 89.1                        & 73.2                      & 29.1                    & \textbf{75.4}                    & 64.9   \\
CLAP+Fusion            & 88.0                        & 75.8                    & 26.3                  &      75.3                 &     64.4  \\
CLAP+K2C Aug.           & \textbf{91.0}                        & \textbf{77.0}                      & \textbf{46.2}                    &         75.3               &   59.7     \\ \hline
SoTA*                  & 82.6\cite{clap-retrieval}                        & 73.2\cite{clap-retrieval}                      & 10.0\cite{wav2clip}                    & 64.1\cite{mbt}                    & \textbf{65.6}\cite{passt}   \\ \hline\hline
\end{tabular}}
\caption{The zero-shot (ZS.) and supervised (SV.) audio classification results. The SoTA of each dataset/setting is denoted by the reference after the number.}
\vspace{-0.6cm}
\label{tab:exp-ac}
\end{table}

\vspace{-0.2cm}
\subsection{Zero-shot and Supervised Audio Classification}

\noindent \textbf{Zero-shot Audio Classification} To study the model generalization and robustness, we conduct zero-shot audio classification experiments on three top-performing models in previous experiments. We evaluate models on three audio classification dataset, namely ESC-50~\cite{esc50}, VGGSound~\cite{vggsound}, and Urbansound8K (US8K)~\cite{us8k}. We use \textbf{top-1 accuracy} as the metric. We classify audio by performing audio-to-text retrieval with each text corresponds to the text prompt converted from class label via
``This a sound of \texttt{label}.". We noticed a dataset overlap between our training data and the zero-shot dataset we are evaluating on. We \textbf{excluded all the overlap samples} and perform zero-shot evaluation on the whole remaining dataset.

\vspace{0.1cm}
\noindent \textbf{Supervised Audio Classification}
We perform supervised audio classification by fine-tuning the audio encoder on FSD50K~\cite{fsd50k} and VGGSounddatasets. We do not conduct this experiment on ESC-50 and Urbansound8K because the potential data leakage issue in those dataset will makes the results incomparable with the previous methods. Specially, \textbf{mAP} is used as the metric to evaluate FSD50K.

As shown in the in Table~\ref{tab:exp-ac}, our models achieves new SoTAs of zero-shot audio classification across all three datasets, demonstrating the high generalization ability of our model to unseen data. Keyword-to-Caption augmentation increases the performance of VGGsound and US8K a lot as it adds more text captions to ``enrich" the text embedding space. Feature fusion not only enables the model to handle variable-length input, but also achieves better performance than previous models. Our best supervised audio classification result outperforms the current state-of-the-art on VGGSound dataset and is close to state-of-the-art on FSD50K dataset. The results verify that the proposed model also learns efficient audio representation during contrastive learning paradigm. 

\vspace{-0.3cm}
\section{Conclusion and Future Work}

In this paper, we propose a large-scale audio-text dataset and improvements on current language-audio contrastive learning paradigm. We show that LAION-Audio-630, AudioSet with keyword-to-caption augmentation, and feature fusion effectively leads to better audio understanding, task performance, and enables effective learnings on variable-length data. Future works include collecting even larger dataset on training, applying representations into more downstream tasks such as audio synthesis and separation.


\bibliographystyle{IEEEbib}
\bibliography{refs}

\clearpage

\appendix
\onecolumn
\section{Appendix}

\section{Acknowledgement}
Yusong Wu, Ke Chen, Tianyu Zhang are opensource contributors to LAION projects. Our codebase is build on following open-source projects: \texttt{PANN}\footnote{\url{https://github.com/qiuqiangkong/audioset_tagging_cnn}}, \texttt{HTSAT}\footnote{\url{https://github.com/RetroCirce/HTS-Audio-Transformer}}, \texttt{open\_clip}\footnote{\url{https://github.com/mlfoundations/open_clip}}, \texttt{PyTorch}\footnote{\url{https://pytorch.org/}}. We would like to thank the support of computation infrastructure from LAION, Stability.ai and Summit cluster from Oak Ridge National Laboratory. This project was proposed in IFT-6167 at University of Monreal and Mila given by Prof. Irina Rish. We would like to thank the support from Christoph Schuhmann, Richard Vencu,, Romain Beaumon, as this project would not be possible without them. We would like to thank the Institute for Research and Coordination in Acoustics and Music (IRCAM)\footnote{\url{https://www.ircam.fr/}} and Project \texttt{REACH}\footnote{\url{https://www.ircam.fr/projects/pages/reach-project}}: Raising Co-creativity in Cyber-Human Musicianship for supporting this project. We would like to thank all the community contributors for contributing the collection of LAION-630k dataset. Those community contributors (Discord ids) include but not limited to: @marianna13\#7139, @Chr0my\#0173, @PiEquals4\#1909, @Yuchen Hui\#8574, @Antoniooooo\#4758, @IYWO\#9072, krishna\#1648, @dicknascarsixtynine\#3885, and @turian\#1607. We would like to thank Xinhao Mei for explaining and helping on retrieval metrics.

\section{Details of Evaluating Retrieval Performance}

In this study, the primary focus is on assessing the efficacy of the models in terms of retrieval performance, utilizing metrics such as R@1, R@5, R@10 and Mean Average Precision (mAP). The Clotho and AudioCaps datasets, in particular, are characterized by the presence of five text ground-truths per audio sample. Therefore, in evaluating the retrieval performance on these datasets, we adopt the same metrics as used in previous studies, specifically, those outlined in \cite{ml-act,mmt}\footnote{We implemented the exact evaluation metric in \url{https://github.com/XinhaoMei/audio-text_retrieval/blob/main/tools/utils.py\#L74}}.

For text-to-audio retrieval, we treat each text from an audio as independent test sample, and calculate the average of text-to-audio retrieval metrics on test samples that are five times the size of test set. In evaluating audio-to-text recall, the recall for each audio is calculated by taking the best audio-to-text retrieval result from the five text ground-truths. Additionally, audio-to-text Mean Average Precision (mAP) is calculated as $mAP@10 = \frac{1}{R} \sum_{r=1}^{10} (P(r) * rel(r))$, where $P(r)$ represents the precision at recall level $r$, and $rel(r)$ is a binary indicator of whether the text at recall level $r$ is relevant or not.

In the case of other datasets, such as Freesound, in which there is only one text associated with each audio sample, the recall and mean average precision (mAP) are measured in the standard manner.

\section{Details of LAION-AUDIO-630K}
Regarding the section 2.1 and section 2.2 of the paper:
\begin{itemize}[leftmargin=*]
    \item We list the specifications of website/sources from which we collect the audio samples and text captions for LAION-Audio-630K in Table~\ref{tab:Laion630k-details}.
    \item We list the details of three datasets in Table~\ref{tab:training-datasets}. We use the combination of them to train the model in the section 4 of the submission.
    \item Regarding the section 3.4 of the paper, we present the distribution of audio length on Epidemic Sound and Freesound\cite{font2013freesound}, as parts of LAION-Audio-630K, to demonstrate the existence of variable-length problem in audio data processing and model training.
\end{itemize}
\label{sec:Details of LAION-630k}
\begin{table}[h]
\centering
\begin{tabular}{lccc}
\toprule
Data Source   & Number of Samples   & \multicolumn{1}{l}{Duration}  & Data Type  \\ 
\midrule
\href{https://sound-effects.bbcrewind.co.uk}{\color{blue}BBC sound effects} & 15973 & 463.48hrs & 1 caption per audio, audio \\
\href{https://www.freetousesounds.com/product/all-in-one-sound-library-bundle}{\color{blue}Free To Use Sounds} & 6370 & 175.73hrs & Filename as caption, audio \\
\href{https://sonniss.com/gameaudiogdc/}{\color{blue}Sonniss Game effects} & 5049 & 84.6hrs & Filename as caption, audio \\
\href{https://www.wesoundeffects.com/}{\color{blue}We Sound Effects} & 488 & 12.00hrs & Filename as caption, audio \\
\href{https://www.paramountmotion.com/odeon-sound-effects}{\color{blue}Paramount Motion Sound Effects} & 4420 & 19.49hrs & Filename as caption, audio \\
\href{https://audiostock.net/se}{\color{blue}Audiostock} & 10000 & 46.30hrs & 1 caption per audio, audio \\
Freesound \cite{font2013freesound} & 515581 & 3003.38rs & 1-2 captions per audio, audio \\
\href{https://www.epidemicsound.com/sound-effects/}{\color{blue}Epidemic Sound} & 75645 & 220.41hrs & 2 captions per audio, audio \\
\bottomrule
\end{tabular}
\caption{LAION-Audio-630k Datasets}
\label{tab:Laion630k-details}
\end{table}

\begin{table}[h]
\centering
\resizebox{0.8\textwidth}{!}{
\begin{tabular}{lccc}
\hline
Data Source   & Number of Samples   & \multicolumn{1}{l}{Duration}  & Data Type  \\ \hline
\multicolumn{4}{c}{\cellcolor[HTML]{C0C0C0}\textit{\textbf{AudioCaps + Clotho}}} \\
AudioCaps     &   49274             & 136.87hrs                     & 1 caption per audio, audio \\
Clotho        &   3839             & 23.99hrs                      & 5 captions per audio, audio \\ 
\multicolumn{4}{c}{\cellcolor[HTML]{C0C0C0}\textit{\textbf{LAION-Audio-630K}}}   \\
BBC sound effects & 15973             & 463.48hrs                     & 1 caption per audio, audio \\
Episodesound      & 75645             & 220.41hrs                     & 2 captions per audio, audio \\
freesound     &   414127            & 2528.15hrs                    & 1-2 captions per audio, audio \\
Free To Use Sounds & 6370             & 175.73hrs                     & Filename as caption, audio \\
Sonniss Game effects & 5049             & 84.6hrs                      & Filename as caption, audio \\
We Sound Effects & 488                & 12.00hrs                     & Filename as caption, audio \\
Paramount Motion Sound Effects & 4420             & 19.49hrs                     & Filename as caption, audio \\
Audiostock    &   10000             & 46.30hrs                      & 1 caption per audio, audio \\
&&&\\
FSD50K        &   36796             & 70.39hrs                      & 1 caption per audio, audio \\
MACS          &   3537              & 9.825hrs                      & Several (2$\sim$) captions per audio, audio \\
Wavtext5K     &   4072              & 23.2hrs                       & 1 caption per audio, audio \\
\multicolumn{4}{c}{\cellcolor[HTML]{C0C0C0}\textit{\textbf{AudioSet}}}           \\
AudioSet      &    1912024          & 463.48hrs                     & 2 captions per audio,audio  \\ \hline
\end{tabular}}
\caption{Training Datasets}
\label{tab:training-datasets}
\end{table}

\begin{figure}[t]
    \centering
     \begin{subfigure}[b]{0.39\textwidth}
        \centering
        \includegraphics[width=\textwidth]{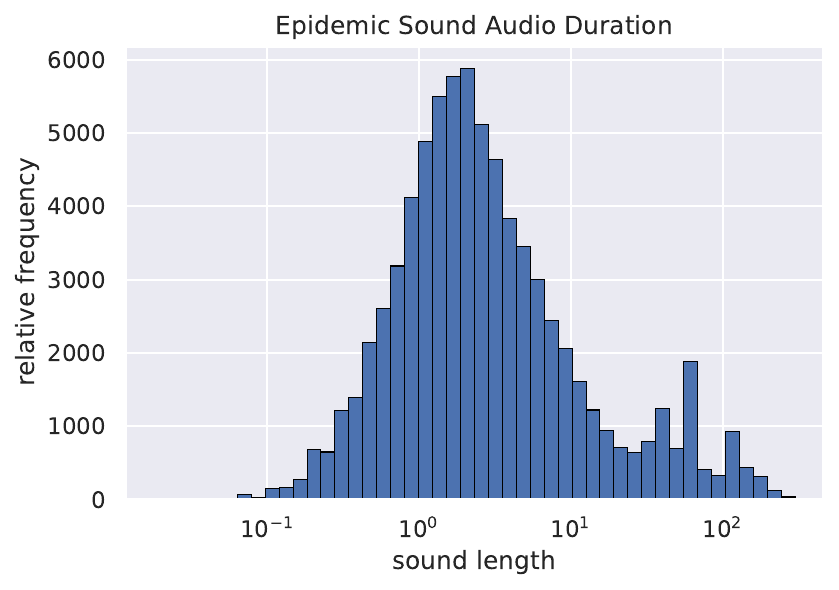}
        \caption{The audio length distribution of Epidemic Sound}
        \label{fig:first}
    \end{subfigure}
    \hspace{2em}
    \begin{subfigure}[b]{0.4\textwidth}
        \centering
        \includegraphics[width=\textwidth]{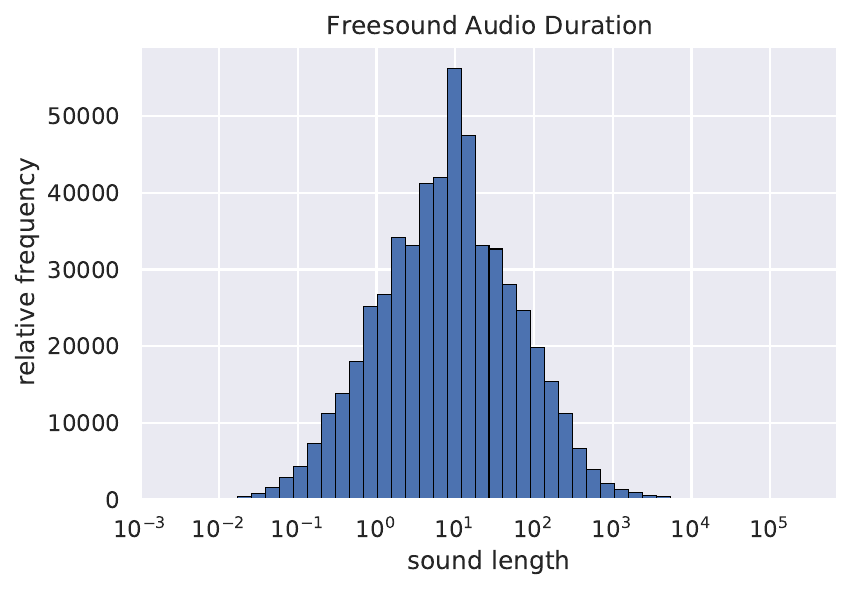}
        \caption{The audio length distribution of Freesound}
        \label{fig:second}
    \end{subfigure}    
\caption{The audio length distribution of Epidemic Sound and Freesound.}
\label{fig:audio-length-distribution}
\end{figure}

\subsection{Freesound Dataset}

The samples in Freesound dataset are collected from Freesound~\cite{font2013freesound}. All audio clips from Freesound are released under Creative Commons (CC) licenses, while each clip has its own license as defined by the clip uploader in Freesound, some of them requiring attribution to their original authors and some forbidding further commercial reuse. Specifically, here is the statistics about licenses of audio clips involved in LAION-Audio-630K:

\begin{itemize}
    \item CC-BY: 196884
    \item CC-BY-NC: 63693
    \item CC0: 270843
    \item CC Sampling+: 11556
\end{itemize}

We listed the licenses for each sample in our dataset release page\footnote{\url{https://github.com/LAION-AI/audio-dataset/tree/main/laion-audio-630k}}.

\section{Attentional Feature Fusion}
Regarding the section 3.4 of the paper, we demonstrate the ``attentional feature fusion" architecture, a two-branch CNN network, to show how we combine the global information and the local information of input audios together.

\begin{figure}[h]
    \centering
    \includegraphics[width=0.32\columnwidth]{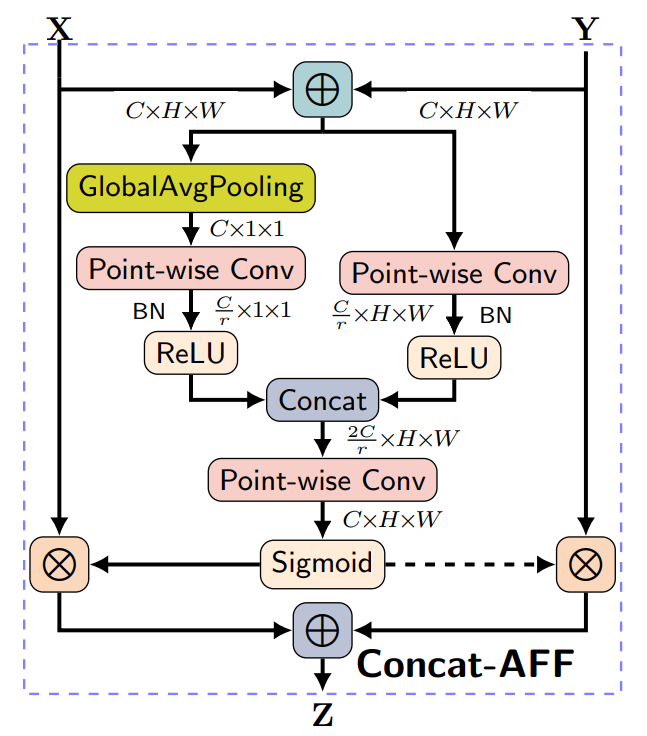}
    \caption{The attentional feature fusion architecture from \cite{aff}.}
    \label{fig:aff}
\end{figure}

As shown in Figure~\ref{fig:aff}, the fusion architecture accepts two inputs: $X$ is the global information ($X_{global}^a$), and $Y$ is the merged local information ($X_{local}^a$) Two inputs are sent to two CNN networks to generate the coefficient, then $X$ and $Y$ are added by this coefficient.

\section{Additional Experiment of feature fusion on Freesound Dataset}
Regarding to the section 4.2 of the paper, to further evaluate the efficacy of feature fusion, apart from AudioCaps and Clotho datasets, we further evaluate our model on Freesound evaluation set, which contains more than 10-sec audio samples (similar to Clotho dataset).

The result is shown in Table~\ref{tab:freesound-map}, the notation is the same as the Table~\ref{tab:exp-t2a-retrieval} in our submission paper.
 The performance on Freesound dataset shares a similar trend with that on Clotho dataset:
\begin{itemize}[leftmargin=*]
    \item  the performance trained on ``AudioCaps + Clotho + LA." is better than that trained on ``AudioCaps + Clotho + LA. + AudioSet". As demonstrate in the section 4.2, similar to Clotho, the Freesound dataset contains audio samples that are different from AudioSet, adding the AudioSet into the training will move the model's distribution out of general audio data to AudioSet-like audio data, such decreasing the performance.
    \item the performance with feature fusion is better than that without feature fusion, as the Freesound dataset contains the samples larger than 10-secs, which is the same to Clotho dataset. Their performance trend are similar.
\end{itemize}
From the above experiment, we can further conclude that the feature fusion can improve the performance of text-to-audio task (i.e., generate better audio representations) on the variable-length audio samples.

\begin{table}[h]
\centering
\begin{tabular}{lccc}
\hline
\multicolumn{1}{c}{\multirow{2}{*}{Model}} & \multirow{2}{*}{Training Set}                  & \multicolumn{2}{c}{Freesound (mAP@10)} \\ \cline{3-4} 
\multicolumn{1}{c}{}                       &                                                & A$\rightarrow$T  & T$\rightarrow$A \\ \hline
HTSAT-RoBERTa                              & AudioCaps + Clotho + LA.                       & 25.9               & 24.5              \\
HTSAT-RoBERTa (fusion)                     & AudioCaps + Clotho + LA.                       & \textbf{26.4}      & \textbf{24.9}     \\
HTSAT-RoBERTa                              & AudioCaps + Clotho + LA. + AudioSet (K2C Aug.) & 22.9               & 21.8              \\
HTSAT-RoBERTa (fusion)                     & AudioCaps + Clotho + LA. + AudioSet (K2C Aug.) & 24.6               & 22.9              \\ \hline
\end{tabular}
\caption{The text-to-audio retrieval performance on Freesound evaluation set.}
\label{tab:freesound-map}
\end{table}

\section{Examples of Keyword-to-Caption Augmentation}
\begin{figure}[h]
    \centering
    \includegraphics[width=0.75\columnwidth]{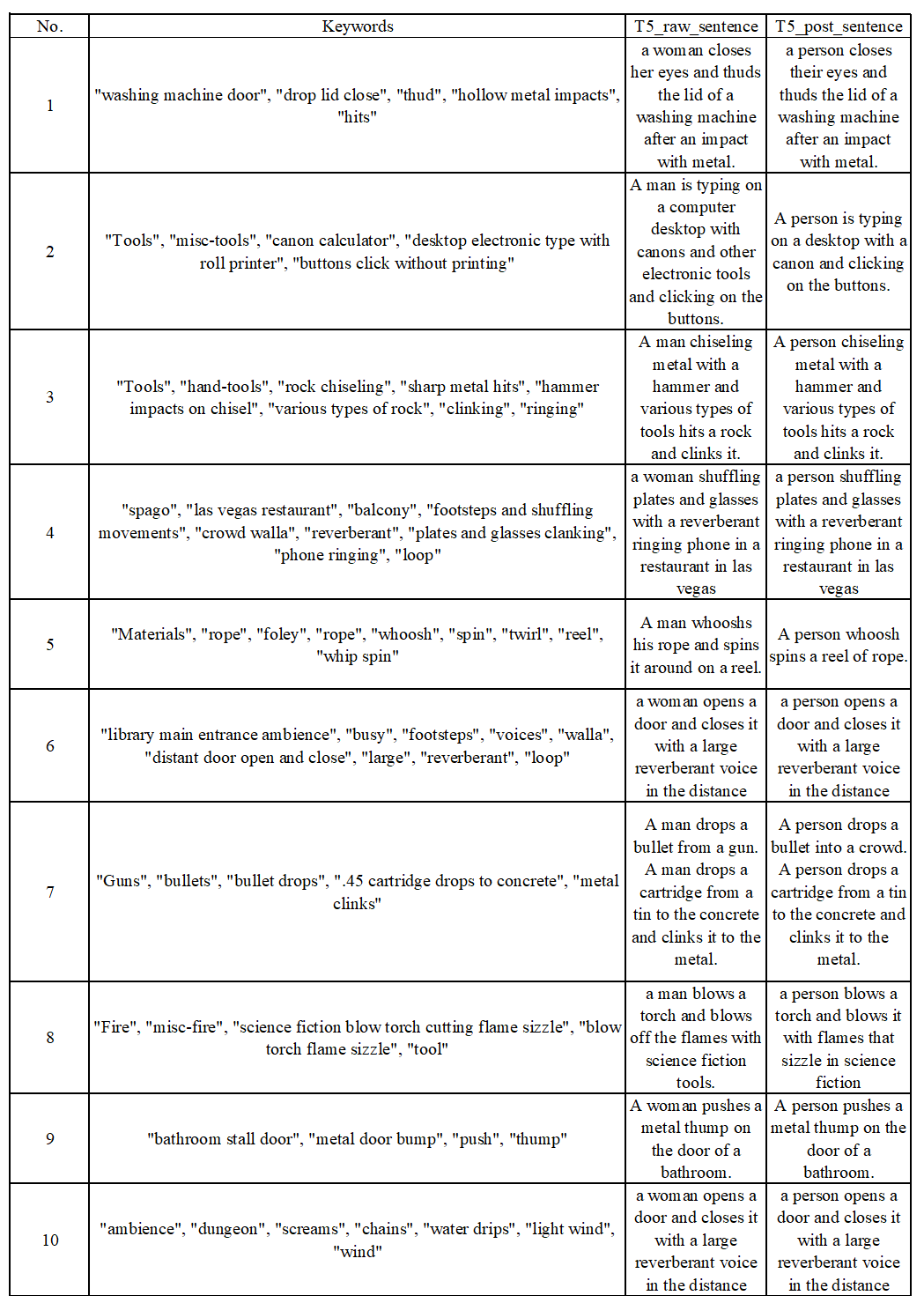}
    \caption{Examples of keyword-to-caption augmentation from AudioSet labels and the de-biased version for the model training.}
    \label{tab:k2c-example}
\end{figure}

Regarding the section 3.5 of the paper, we show some examples of keyword-to-caption by T5 model \cite{t5model}\footnote{We use the T5 model provided in \url{https://github.com/gagan3012/keytotext}.} from AudioSet labels in the below Table~\ref{tab:k2c-example}. And the de-biased version for the model training.

Additionally, when applying keyword to caption, we excluded samples shorter than 2 seconds, as we found in such case the audio is merely a single event, thus matching poorly with the caption generated. When using keyword to caption in training dataset including audioset, we use only the captions generated by keyword to caption and exclude the captions generated by template.

\clearpage
\section{Experiment Settings on Data Exclusion}
Regarding the section 4.3 of the paper, we excluded all the overlap samples and perform zero-shot evaluation on the whole remaining dataset. The below table~\ref{tab:data-exclude} shows the detail of it.

\begin{table}[h]
\centering
    \begin{tabular}{m{3cm}<{\centering}m{3cm}<{\centering}m{3.1cm}<{\centering}}
    \toprule
        Datasource A & Datasource B & Number of samples from Datasource A that are also in Datasource B  \\
        \midrule
        ESC50-all & Clotho-train &  94  \\
        ESC50-all & Clotho-valid &  27  \\ 
        ESC50-all & Clotho-test & 34 \\ 
        ~ & ~ & ~ \\ 
        ESC50-all & FSD50K-train & 399 \\ 
        ESC50-all & FSD50K-valid & 60 \\ 
        ESC50-all & FSD50K-test & 171 \\ 
        ~ & ~ & ~ \\ 
        USD8K-all & Clotho-train & 411 \\ 
        USD8K-all & Clotho-valid & 150 \\ 
        USD8K-all & Clotho-test & 209 \\ 
        ~ & ~ & ~ \\ 
        USD8K-all & FSD50K-train & 697 \\ 
        USD8K-all & FSD50K-valid & 180 \\ 
        USD8K-all & FSD50K-test & 341 \\ 
        ~ & ~ & ~ \\ 
        Clotho-test & FSD50K-train & 54 \\ 
        Clotho-test & FSD50K-valid & 15 \\ 
        Clotho-test & FSD50K-test & 33 \\ 
        ~ & ~ & ~ \\ 
        FSD50K-test & Clotho-train & 137 \\ 
        FSD50K-test & Clotho-valid & 31 \\ 
        FSD50K-test & Clotho-test & 33 \\ 
        ~ & ~ & ~ \\ 
        Clotho-valid & FSD50K-train & 53 \\ 
        Clotho-valid & FSD50K-valid & 10 \\ 
        ~ & ~ & ~ \\ 
        FSD50K-valid & Clotho-train & 38 \\ 
        FSD50K-valid & Clotho-valid & 10 \\ 
        ~ & ~ & ~ \\ 
        Audiocaps-test & Audioset-unbalanced-train & 4875 \\ 
        Audiocaps-test & Audioset-balanced-train & 0 \\ 
        ~ & ~ & ~ \\ 
        audioset-test & audiocaps-train & 0 \\ 
        audioset-test & audiocaps-valid & 0 \\ 
        \bottomrule
    \end{tabular}
\caption{ The overlaps between the training data and the zero-shot evaluation data, we excluded all these overlaps from the evalation sets to calculate the audio classification metrics.}
\label{tab:data-exclude}
\end{table}

\end{document}